\documentclass[11pt]{article}

\usepackage[margin=1in]{geometry}
\usepackage[T1]{fontenc}
\usepackage{lmodern}
\usepackage{amsmath}
\usepackage{amssymb}
\usepackage{mathtools}
\usepackage{microtype}
\usepackage[round,authoryear]{natbib}
\usepackage{hyperref}

\hypersetup{
  colorlinks=true,
  linkcolor=blue,
  citecolor=blue,
  urlcolor=blue
}

\newcommand{\doi}[1]{\href{https://doi.org/#1}{doi:#1}}
\newcommand{\PrO}{\Pr\nolimits_O}
\newcommand{\PrR}{\Pr\nolimits_R}

\title{The Future-Record Model of Everettian Probability}
\author{Jonathan Baxter}
\date{September 3, 2026}

\begin{document}

\maketitle

\begin{abstract}
I recently showed how objective Everettian probability can be defined over
the records borne by an observer's later continuations, without unique
actualization or additional premeasurement subjects \citep{Baxter2026}.
Here I extend that model to all decoherent physical record states, including
records that support no such continuation. The observer-indexed model is then
a restriction of an exhaustive physical-record model.
\end{abstract}

\section{The observer-indexed future-record model}
\label{sec:observer-model}

In \citet{Baxter2026}, I proposed an Everettian probability model whose sample
points are a present observer's later continuations and whose random variable
returns each continuation's record. This note extends that model to all
decoherent physical record states, including those that support no such
continuation. Its only additional interpretive assumption is that a record's
actuality and objective probability do not depend on its being borne by an
observer.

Consider an idealized measurement in which every record value with nonzero
Born weight is borne by a later continuation of the premeasurement observer
\(O_{\mathrm{pre}}\). Let \(I\) be the finite set of record values and let
\(O_i\) be the later observer bearing record \(i\). The one-to-many
continuation relation is
\begin{equation}
  O_{\mathrm{pre}}
  \longrightarrow
  \{O_i:i\in I\}.
  \label{eq:observer-continuations}
\end{equation}
The observer-indexed sample space is the set of later observers on the
right-hand side:
\begin{equation}
  \Omega_O=\{O_i:i\in I\}.
  \label{eq:observer-space}
\end{equation}
The future-record variable is
\begin{equation}
  X_O:\Omega_O\longrightarrow I,
  \qquad X_O(O_i)=i.
  \label{eq:observer-variable}
\end{equation}
If \(w_i\) is the normalized Born weight of record \(i\), the observer-level
Born probability postulate is
\begin{equation}
  \PrO(X_O=i)=w_i,
  \qquad \sum_{i\in I}w_i=1.
  \label{eq:observer-law}
\end{equation}

Every \(O_i\) is actual, but each bears only record \(i\). Thus \(X_O=i\)
concerns the local record at \(O_i\), not whether that observer exists
globally. This is the observer model's interpretive work: it identifies
mutually exclusive local record values as the objects of prospective
probability without selecting a unique later observer. The detailed argument
from delayed revelation and one-to-many continuation is given in
\citet{Baxter2026}. The probability interpretation remains an explicit
postulate; it is not derived from the continuation relation or Born measure
alone.

However, the model assumes that every record alternative supports a later
continuation of \(O_{\mathrm{pre}}\). Physical experiments need not satisfy
that assumption.

\section{The physical-record extension}
\label{sec:record-model}

As just noted, the observer model need not exhaust the physical outcomes.
Lewis and Baker illustrate this with a quantum measurement that controls a lethal device,
leaving a later observer in one outcome and none in the other. Lewis presents
the case as a problem for Everettian probability and decision
\citep{Lewis2007}. Baker isolates the formal problem: if outcomes are
identified with future observers, the lethal outcome is missing despite its
nonzero Born weight \citep{Baker2007}. He replaces future observers with
classes of branches. Here I make the same essential enlargement at the level
of physical records, without counting individual branches.

Let \(R_{\mathrm{pre}}\) denote the premeasurement physical state. Under
unitary evolution, the measurement interaction and environmental decoherence
produce a normalized postmeasurement state \(\lvert\Psi\rangle\) with record
alternatives\footnote{The arrow marks the direction of stable record
formation, not a fundamental irreversibility in the unitary dynamics.}
\begin{equation}
  R_{\mathrm{pre}}
  \longrightarrow
  \{R_i:i\in I\}.
  \label{eq:record-alternatives}
\end{equation}
Here \(I\) is an exhaustive finite set of macroscopic record values with
nonzero Born weight, and \(R_i\) is the complete coarse-grained physical
alternative bearing record \(i\), including any apparatus, environmental,
and observer states within that sector. Let \(\Pi_i\) project onto the
corresponding sector, with
\begin{equation}
  \Pi_i\Pi_j=0\quad(i\ne j),
  \qquad
  \sum_{i\in I}\langle\Psi\rvert\Pi_i\lvert\Psi\rangle=1.
  \label{eq:exhaustive-projectors}
\end{equation}
The record sample space is
\begin{equation}
  \Omega_R=\{R_i:i\in I\},
  \label{eq:record-space}
\end{equation}
and the record variable is
\begin{equation}
  X_R:\Omega_R\longrightarrow I,
  \qquad X_R(R_i)=i.
  \label{eq:record-variable}
\end{equation}

The Born measure of a record alternative is
\begin{equation}
  \mu_{\Psi}(R_i)
  =\langle\Psi\rvert\Pi_i\lvert\Psi\rangle
  =w_i.
  \label{eq:record-measure}
\end{equation}
The observerless Born probability postulate interprets this measure as
objective probability:
\begin{equation}
  \boxed{
  \PrR(X_R=i)
  =\mu_{\Psi}(R_i)
  =w_i.}
  \label{eq:record-law}
\end{equation}

``Observerless'' describes the sample-space definition, not the contents of
every sample point. Some \(R_i\) include observers and others do not. A record
is a stable physical correlation encoded in an apparatus or environment; it
need not be cognitively available. ``Future'' is relative to the physical
preparation, not necessarily to an observer. The extension assumes that a
physical record's actuality and objective probability do not depend on its
being borne by an observer. A view requiring a subject for a record to be
actual will reject that assumption. The formal construction cannot settle the
disagreement.

Every record sector with nonzero Born weight occurs in the Everettian state,
but \(X_R\) has exactly one value at each \(R_i\). Equation~\eqref{eq:record-law}
is therefore the same axiomatic step as Equation~\eqref{eq:observer-law}, now
defined on an exhaustive physical domain.

Each \(R_i\) is the coarse-grained record alternative associated with
\(\Pi_i\), not an individually counted branch. A finer orthogonal
decomposition preserving record \(i\) introduces no new value of \(X_R\) and
leaves its total Born weight \(w_i\) unchanged.

\section{The observer model as a special case}
\label{sec:special-case}

Suppose that \(R_{\mathrm{pre}}\) supports \(O_{\mathrm{pre}}\) and that every
\(R_i\) supports the corresponding later observer \(O_i\). Define the map
from each observer continuation to its supporting physical continuation:
\begin{equation}
  f:\Omega_O\longrightarrow\Omega_R,
  \qquad f(O_i)=R_i.
  \label{eq:model-map}
\end{equation}
Then \(f\) is a bijection at the coarse-grained record level and
\begin{equation}
  X_O=X_R\circ f.
  \label{eq:variables-commute}
\end{equation}
It follows immediately that, for every \(A\subseteq I\),
\begin{equation}
  \PrO(X_O\in A)=\PrR(X_R\in A).
  \label{eq:model-equivalence}
\end{equation}
The two models therefore have the same record events and probability
distribution whenever the observer space is exhaustive.

If only the outcomes in a nonempty subset \(J\subsetneq I\) support later
observers continuing \(O_{\mathrm{pre}}\), let
\begin{equation}
  \Omega_O^J=\{O_i:i\in J\},
  \qquad
  C_J=\{R_i:i\in J\}\subseteq\Omega_R.
\end{equation}
The restricted observer domain carries total Born weight
\(\sum_{i\in J}w_i<1\), so it is not an exhaustive probability model.
Renormalizing it would produce
\begin{equation}
  \frac{w_i}{\sum_{j\in J}w_j}
  =\PrR(X_R=i\mid C_J),
  \qquad i\in J,
  \label{eq:conditional-restriction}
\end{equation}
but this answers which record is borne conditional on an outcome supporting
a later continuation of \(O_{\mathrm{pre}}\). It does not replace the
unconditional record law.

Although the observer domain is formally a restriction of the record domain,
the original observer model provides the
interpretive starting point \citep{Baxter2026}. Delayed revelation shows that
an observer can be uncertain about their local record while knowing that every
record exists globally. This identifies the physical record value, rather than
global existence, as the target of uncertainty. Under the observer-independence
assumption, the same law extends to every physical record state. In the reverse
direction, the observer domain and variable are recovered by restricting the
record model to alternatives supporting later continuations of
\(O_{\mathrm{pre}}\). When that restriction is exhaustive,
Equation~\eqref{eq:model-equivalence} also recovers the observer-level
probability law.

Following the use of record likelihoods in \citet{Baxter2026}, I take
\(w_i=\PrR(X_R=i)\) as the likelihood of the record observed by \(O_i\).
Restricting the domain to later observers does not by itself justify replacing
these likelihoods with probabilities conditioned on their existence.

\section{Applications and limits}

\subsection{A lethal measurement}

Apply the record model to a measurement whose survival outcome has Born
weight \(p\) and whose lethal outcome has weight \(1-p\). After decoherence,
write the relevant state schematically as
\begin{equation}
  \lvert\Psi_{\mathrm{post}}\rangle
  =\sqrt{p}\,\lvert R_S\rangle
   +\sqrt{1-p}\,\lvert R_D\rangle.
  \label{eq:lethal-state}
\end{equation}
Both \(R_S\) and \(R_D\) are physical continuations of
\(R_{\mathrm{pre}}\). Only \(R_S\) supports a later continuation \(O_S\) of
\(O_{\mathrm{pre}}\); \(R_D\) records \(O_{\mathrm{pre}}\)'s death.

The observerless sample space is
\begin{equation}
  \Omega_R=\{R_S,R_D\},
\end{equation}
and Equation~\eqref{eq:record-law} gives
\begin{equation}
  \PrR(X_R=S)=p,
  \qquad
  \PrR(X_R=D)=1-p.
  \label{eq:lethal-probabilities}
\end{equation}
The observer-indexed domain therefore contains only \(O_S\) and carries total
Born weight \(p\). Renormalizing this one-point restriction would trivially give
\(\PrR(X_R=S\mid C_{\{S\}})=1\). That is a conditional probability given a
continuation-supporting outcome. Renormalization alone does not justify using
it as the likelihood of the physical record \(S\). The exhaustive record law
assigns that event probability \(p\), which is the likelihood used here.

This accepts Baker's central point that future observers do not exhaust the
physical outcomes. It uses record alternatives rather than branch classes and
retains the observer domain as a derived restriction. How
\(O_{\mathrm{pre}}\) should value the two alternatives is a separate question,
discussed by \citet{Baker2007}.

\subsection{Unread records and observer number}

An automated experiment can leave stable physical records even if no observer
ever reads them. The observer model then has no application, but the record
model remains well defined. Likewise, if different record outcomes contain
different numbers of observers, the Born probability of \(R_i\) remains
\(w_i\); it is not multiplied by their number. Sampling or locating an
observer among those outcomes requires a further rule, not supplied by the
physical Born law.

These consequences should not be confused with a general solution to
anthropic probability or rational choice in lethal experiments. The model
distinguishes physical chance from observer-conditioned probability. It does
not by itself determine how an agent should value outcomes containing
different observers, nor does it derive a principle for sampling among them.

\section{Conclusion}

The observer model establishes the interpretation of Everettian probability
where the familiar objection is strongest: every later observer is actual,
but each bears only one local record. Once that model identifies the physical
record value as what probability concerns, extending the law to every
decoherent physical record alternative is immediate, given the
observer-independence assumption.

When every record supports a later continuation of \(O_{\mathrm{pre}}\), the
observer model is coextensive with the record model. When only some do, the
observer domain is a nonexhaustive restriction, not a separate unconditional
probability model.
The record model covers lethal outcomes, unread records, and outcomes with
different observer populations without making physical probability depend on
the existence or number of observers.

\end{document}